# Direct and absolute temperature mapping and heat transfer measurements in diode-end-pumped Yb:YAG

**Sébastien Chénais, Sébastien Forget, Frédéric Druon, François Balembois and Patrick Georges**

Laboratoire Charles Fabry de l'Institut d'Optique, Centre scientifique bât. 503, 91403 Orsay cedex, France. Tel : +331 6935 8792; fax : +331 6935 8807 ;

Abstract : We report direct and absolute temperature measurements in a diode-end-pumped Yb:YAG crystal, using a calibrated infrared camera, with a 60-µm spatial resolution. The heat transfer coefficient has been measured, for the first time to our knowledge, with four different types of thermal contact (H = 0.25, 0.28, 0.9 and 2.0 for bare contact, graphite layer, indium foil and heat sink grease respectively). The dynamics of thermal effects is also presented.

## 1. Introduction

The achievement of high power laser systems with high beam quality is largely compromised by thermal effects in the gain medium, which are responsible for thermal lensing, depolarisation losses, and ultimately fracture [1]. The most widespread way to investigate thermal effects in laser crystals is to use thermo-optical methods, that is thermal lensing or depolarization measurements. However, these methods only yield quantities that are proportional to thermal gradients inside the probed area and are related to some more or less known thermo-optic coefficients. Similarly, the temperatures that are obtained by finite element calculations are only relative temperature distributions, expressed with respect to the rod surface temperature. The latter depends on the boundary conditions and is then very difficult to predict. Direct temperature mapping could consequently be a helpful measurement to understand pump-induced thermal effects. Moreover, it is well known that one of the crucial parameter to reduce heating effects and to avoid fracture is the thermal contact between the crystal and its surrounding mount. Consequently, the knowledge of quantitative information as the heat transfer coefficient is very important for high power laser development.

The aim of this letter is to present experimental absolute temperature maps obtained in the well-known Yb:YAG [2] crystal in end-pumping configuration. Our setup allows spatially resolved analysis of the temperature on the entrance face of the crystal, where temperature reaches generally its maximum value. In the same time, the temporal evolution of the temperature can be observed for each point of the crystal surface. We have also measured for the first time to our knowledge the heat transfer coefficient between the crystal and its surrounding for different types of often used thermal contacts.

## 2. Experimental setup

We present non-contact absolute measurements performed with an infrared imager working in the 8-12 µm spectral range. The experimental setup (figure 1) allows direct measurements of the temperature inside the laser crystal exactly in the end-pumping conditions. The crystal used here was a 2-mm long, 4x4 mm$^2$ square cross section, 8 at. % Yb:YAG crystal. It was AR-coated on its faces (the lateral ones are polished). Its thermal conductivity, which is lower than that of an undoped YAG crystal, was measured to be 0.07 W.cm$^{-1}$.K$^{-1}$ (0.11 W.cm$^{-1}$.K$^{-1}$ for the undoped crystal). The pump source was a high power fiber-coupled diode array (HLU15F200-980 from LIMO GmbH) emitting 13.5 W at 968 nm. The fiber had a core diameter of 200 µm and a numerical aperture of 0.22. The output face was imaged onto the crystal to a 270-µm-diameter spot via two doublets. The crystal absorbed 5.4 watts of pump power in this case. The high spatial resolution of the thermal imager was obtained thanks to a dichroic Zinc selenide plate, High Reflectivity (HR) coated for 960-1080 nm on one face (at 45° angle of incidence), and also coated for High Transmission (HT) in the 8-12 µm spectral range on both faces (coating from *Opticorp Inc.*). An aberration-free germanium objective (focal length 50 mm, N.A. 0.7) was appended close to the ZnSe plate to create the intermediate thermal image. The camera was an AGEMA 570 (*Flir Systems Inc.*) consisting of 240x320 microbolometers working at room temperature. The measured noise equivalent temperature difference (NETD) of the camera is 0.2 °C. The numerical aperture of the whole imaging system in the object medium being around 1, a theoretical spatial resolution of 10 µm could be achieved; however, the resolution is here limited to 60 µm by the size of the pixels. The crystal was clamped in a copper block by its four side faces. In addition, a frictionless copper finger and a set of known weights allowed to apply a given pressure on the top face of the crystal. The heat is evacuated from the copper block by a flow of circulating water.

The key issue of infrared absolute temperature measurements is the correct calibration of the system. Indeed, neither the crystal nor the copper mount has an infrared luminance which equals that of a blackbody at the same temperature. The signal V detected by one pixel for a portion of crystal (or copper mount) at temperature T is:

$$V(T) = G \int_{\Delta \lambda} S_r(\lambda) \left[ Tr_{opt} \left( \varepsilon(T) \frac{dL_{BB}^T}{d\lambda} + L_t \right) + L_r \right] d\lambda \qquad (1)$$

where G is the geometric extent; $S_r(\lambda)$ is the spectral sensitivity; $Tr_{opt}$ is the whole transmission coefficient of the ZnSe plate, Germanium objective and camera optics; $\frac{dL_{BB}^T}{d\lambda}$ is the spectral luminance of a blackbody at temperature T, ε(T) is the emissivity; $L_r$ denotes the infrared luminance of the camera itself (and its close surroundings) which is reflected back into it by the Germanium objective and by the polished surface of the crystal; $L_t$ is the luminance transmitted through the crystal: it is zero in the 8-12 µm range since the crystal is highly opaque in this spectral region. $L_r$ is nonzero and makes polished objects look brighter than blackbodies: if $L_r$ is ignored it leads to overestimation of the temperature around room temperature. Inversely, the emissivity is less than one and makes objects radiate less than a blackbody. Since the parameters ε and $L_r$ are dramatically dependant on the surface quality and flatness, all the visible parts of the heat sink were covered with lustreless black painting. Moreover, the evaluation of all those parameters is not straightforward. We propose to calibrate the whole system as follows: the crystal and the copper mount were heated together to a set of given temperatures using a thermoelectric element, and we then compare with the temperature given by the camera to apply the adequate correction. This careful calibration allows rigorous and absolute measurement of the temperature with a spatial resolution large enough to study with sufficient accuracy the thermal behaviour on the crystal's entrance face.

## 3. Results
Figure 2 shows the temperature map obtained when the crystal is clamped by its four edge faces by bare contact with copper without thermal joint (left part) and with heat sink grease

(right part). In the first case, a clear gap is noticeable between the temperatures of the mount and at the edge of the crystal. The temperature distribution is parabolic inside the pumped region and then experiences a logarithmic decay until the edge of the crystal, in good agreement with the theory in the case of fiber-coupled diode pumping. The temperature difference at the centre of the crystal between the two types of thermal contact shows clearly the importance of the interface.

We consequently studied more in details the heat contact. By analogy with convective transfer, the quality of the contact can be accounted by a heat transfer coefficient H (in W.cm$^{-2}$.K$^{-1}$), defined so that the heat flux through the surface is [3]:

$$-K_c \left\| \vec{\nabla} T \right\|_e = H(T_e - T_m) \qquad (2)$$

where $K_c$ is the thermal conductivity of the crystal (W.cm$^{-1}$.K$^{-1}$), $T_e$ the temperature (and $\left\| \vec{\nabla} T \right\|_e$ the temperature gradient) at the edge of the crystal and $T_m$ the heat sink temperature. The thermal gradient in (2) is considered normal to the surface. An ideal contact ($T_e = T_m$) corresponds to an infinite value of H.

Our system provides a space-resolved temperature mapping of the crystal, with a spatial resolution which is far below the crystal size: it then allows the measurement of H, which was never measured in the case of end-pumping.

By performing a linear fit of the temperature versus position on the points that are closer to the crystal edge, the heat flux can be determined: by applying relation (2), one can then infer the value of H, which is here equal to 0.25 W.cm$^{-2}$.K$^{-1}$ in the case of bare contact. We estimate that the uncertainty on H is about 15%. The order of magnitude obtained is consistent with the values evoked by Carslaw [3] and Koechner [4]. A hot spot can be noticed in figure 2, which betrays the poor contact between the polished face of the crystal and the copper surface. The heat transfer is primarily a question of how much two surfaces are in contact with respect to each other, and we check experimentally that the temperature inside the crystal does not depend on the applied pressure : we did not observe any noticeable variation of the temperature when changing the applied pressure if there is no thermal joint between the crystal and the copper mount. However, the heat transfer is strongly dependant on how the copper surface is positioned with respect to the crystal surface on which it applies. We have then tested three different thermal joints inserted between the crystal and the copper: graphite layer, indium foil and heat sink grease (CT40-5 from Circuitworks). The results are summarized in table 1. Graphite layer (≈ 0.5 mm thick) changes significantly neither the maximum temperature nor the heat transfer coefficient, but it was remarked that the contact was much more uniform than with bare contact: in particular no hot spot appeared any more and the contact was somewhat independent of the applied pressure. It is in contrast not the case with indium foil. For this experiment the crystal was wrapped with a 1-mm thick indium foil. Since Indium is a soft material, the quality of the contact is overly dependant on the applied pressure. The temperature at the center of the pumped region experiences a 7°C decrease while the pressure increased from 1.5 kg/cm$^2$ to 22 kg/cm$^2$. Note that in this case the measured H (always defined by (2)), measured across the surface where the pressure is applied is in reality an "effective" heat coefficient that takes into account the transfer from crystal to indium and then from indium to copper. The most dramatic change in heat transfer coefficient is obtained with heat sink grease (see table 1). The temperature gap drops down to 1°C and H reaches 2 W.cm$^{-2}$.K$^{-1}$. The heat contact is here independent on the applied pressure. However long term pollution caused by the grease diffusion inside the crystal should be considered in this latter case.

The dynamics of the temperature installation was also studied, and the results are featured in figure 3. The temperature is shown versus time at the center of the pumped area ($T_{max}$), at the edge of the pumped area ($T_a$), at the edge of the crystal ($T_e$) and just behind the gap in the mount volume ($T_m$). The difference $T_{max}-T_a$, referred as the thermal gradient inside the pumped area, is responsible for thermal lensing effects. This gradient turns out to establish very quickly, that is with a time constant far below 100 milliseconds (corresponding to the response time of the infrared camera). This behaviour has been confirmed by thermal lensing measurements performed with a Shack-Hartmann wavefront sensor [5]. This is consistent with the fact that the thermal diffusion inside the pumped zone is, within a first approximation, governed by the time constant [6]

$$\tau = \frac{w_p^2}{\kappa}$$

where $\kappa$ is the thermal diffusivity and $w_p$ the radius of the pumped area. For (8-at.%) Yb:YAG, $\tau_1 = 2.6$ ms. In contrast, the temperature at the center and at the edge of the rod have been fitted by a simple exponential growth, and a time constant of $\tau_2 = 1.1$ s was the obtained. This time constant is a function of both crystal size, diffusivity, and cooling parameters [7].

## 4. Conclusion

In summary, we have shown, for the first time to the best of our knowledge, absolute temperature maps in a diode-end-pumped laser crystal (here Yb:YAG) with different types of thermal contact, with a 60-µm spatial resolution, using an infrared camera in the 8-12 µm range. The method is very simple and versatile, and can be also used in the lasing conditions (same crystal, same pump). A careful calibration has been done, in order to take into account not only the finite emissivity of the crystal, but also the reflected luminance coming from the camera itself. A temporal analysis showed that the thermal gradient (responsible for thermal lensing effects) had a far shorter time constant than the temperature field itself, the latter being governed by a time constant in the order of 1 s. We also studied the influence of three thermal joints on the cooling. With respect to bare contact, graphite layer allows gaining uniformity but indium foil allows reaching heat transfer coefficients that are very dependant on the applied pressure, whereas heat sink grease enables an improvement of the contact by a factor of 8, whatever the applied pressure. The method presented in this paper is a very simple way to study qualitatively and quantitatively the thermal contact and could be applied to every type of laser crystal and pumping geometry, including end-pumping.

**Acknowledgements**: The authors gratefully acknowledge Gérard Roger for realizing the mechanical parts and for his helpful advice. We also thank the Ecole Supérieure d'Optique for loaning us the infrared camera and the germanium objective.

**Figure captions**

Table 1 **:** experimental results. $T_{max}$ is the temperature at the center of the pumped region ; $T_e$ is the temperature at the edge of the crystal (averaged on the 4 sides if not symmetrical), and $T_m$ is the copper mount temperature near the crystal.

Figure 1 : experimental setup.

Figure 2 : temperature mapping of the crystal (front view) and lateral cut at y=0 for two different types of thermal contact (direct copper-crystal contact on the left, with grease on the right).

Figure 3 : temporal evolution of the temperature at different position in the crystal : at the center (Tmax), at the edge of the pumped area (Ta), at the edge of the crystal (Te) and in the copper mount (Tm).

Table 1 : experimental results. $T_{max}$ is the temperature at the center of the pumped region ; $T_e$ is the temperature at the edge of the crystal (averaged on the 4 sides if not symmetrical), and $T_m$ is the copper mount temperature near the crystal.

| Contact | **$H$ (W.cm$^{-2}$.K$^{-1}$)** | $T_{max}$ (°C) | $T_e$(°C) | $T_e$-$T_m$ (°C) |
|---|---|---|---|---|
| Bare | **0.25** | 49.8 | 33.5 | 10.7 |
| Graphite layer | **0.28** | 46.5 | 30.5 | 8.7 |
| Indium foil *(applied pressure : 22 kg/cm$^2$)* | **0.9** | 40.0 | 25.1 | 4.9 |
| Heat sink grease | **2.0** | 37.0 | 21.6 | 1.5 |

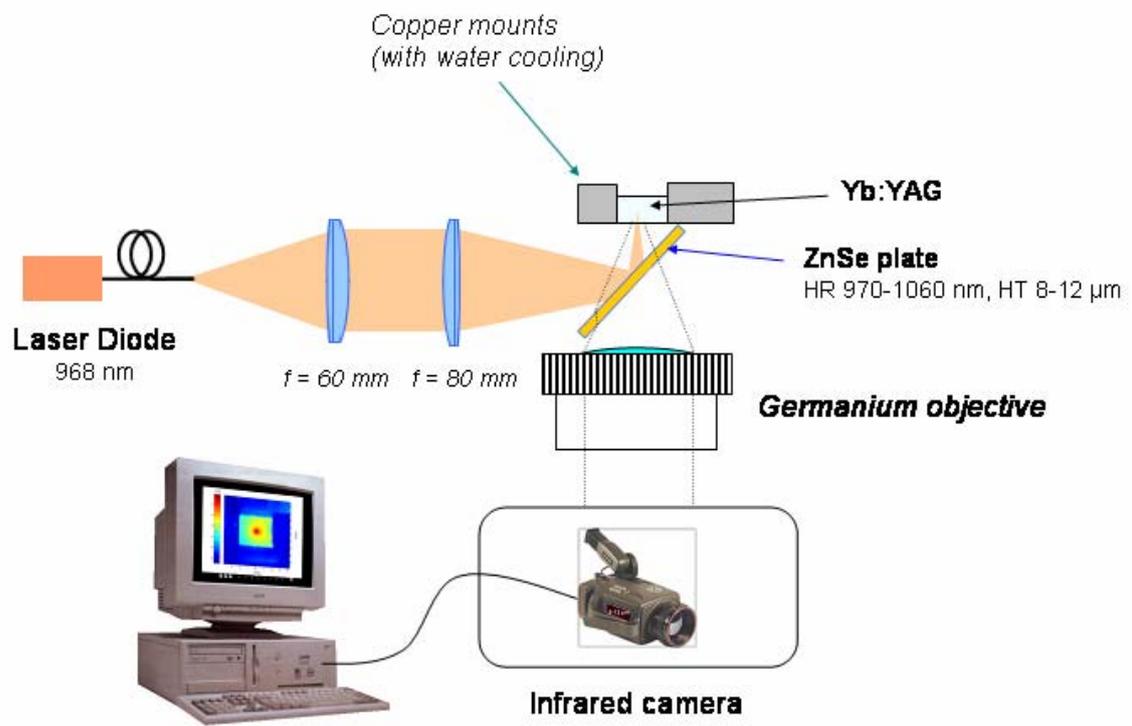

Figure 1 : experimental setup

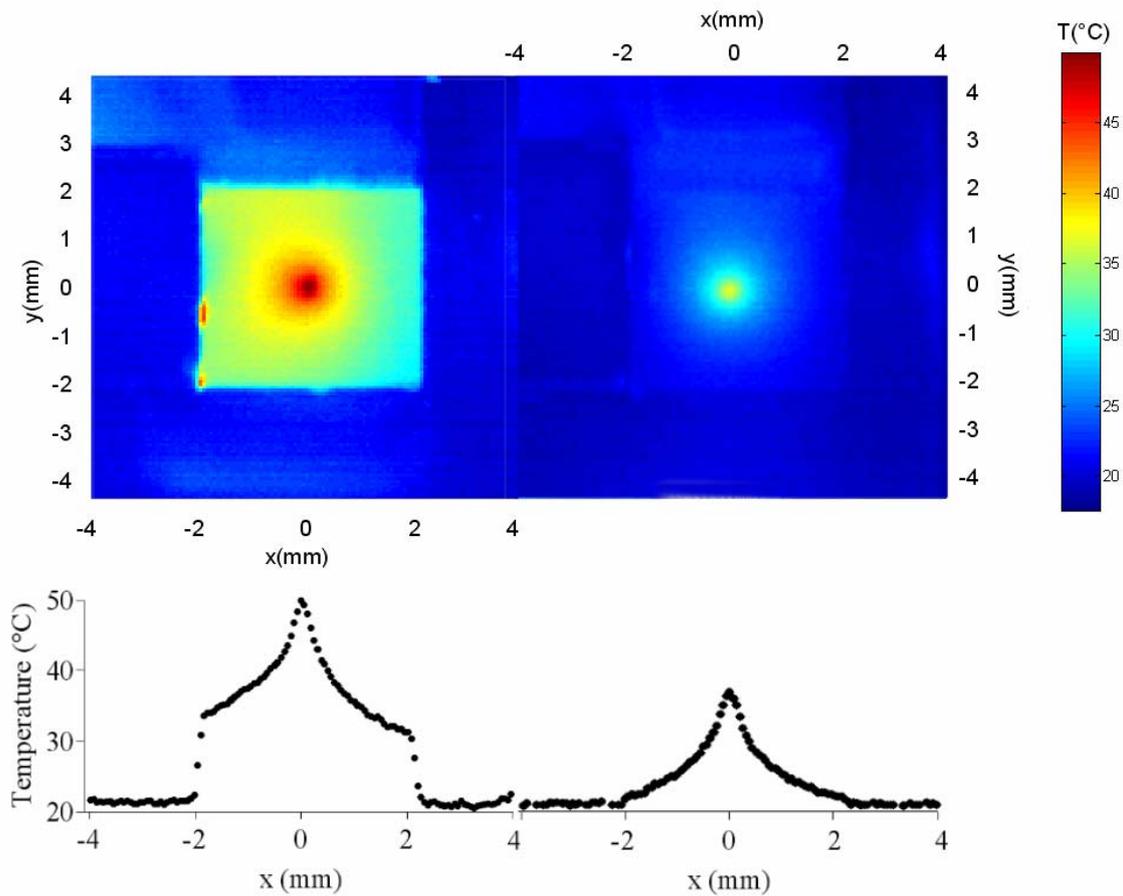

Figure 2 : temperature mapping of the crystal (front view) and lateral cut at y=0 for two different types of thermal contact (direct copper-crystal contact on the left, with grease on the right).

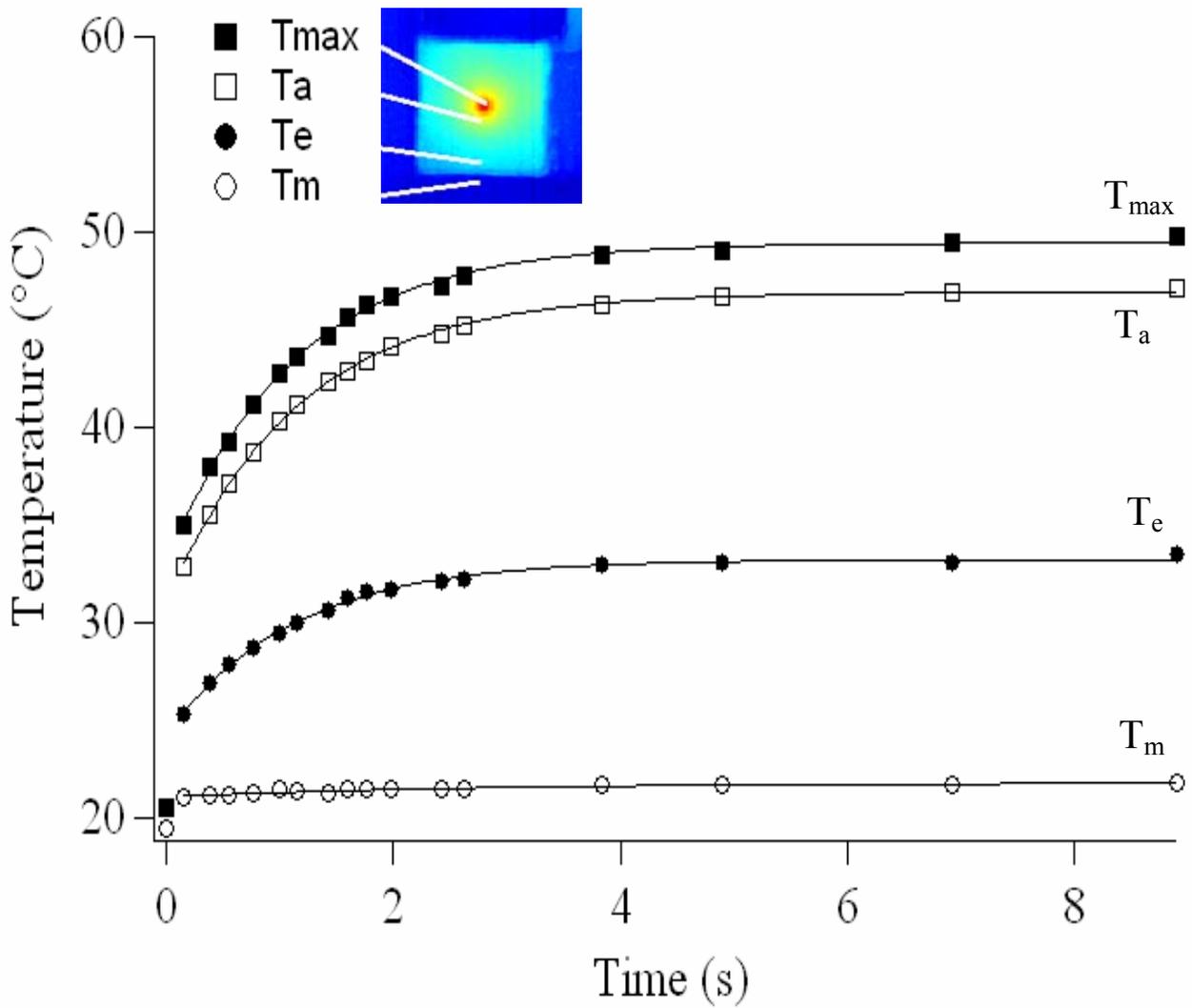

Figure 3 : temporal evolution of the temperature at different position in the crystal : at the center (Tmax), at the edge of the pumped area (Ta), at the edge of the crystal (Te) and in the copper mount (Tm).